# Multi-scale sequence correlations increase proteome structural disorder and promiscuity

Ariel Afek[#], Eugene I. Shakhnovich[##]*, and David B. Lukatsky[#]*

[#]*Department of Chemistry, Ben-Gurion University of the Negev, Beer-Sheva 84105 Israel*
[##]*Department of Chemistry and Chemical Biology, Harvard University*

*Corresponding Authors
D. B. L. Email: lukatsky@bgu.ac.il
Ph.: +972-8-642-8370
Fax: +972-8-647-2943

E. I. S. Email: shakhn@fas.harvard.edu
Ph.: +1-617-495-4130
Fax: +1-617-384-9228

## Abstract

Numerous experiments demonstrate a high level of promiscuity and structural disorder in organismal proteomes. Here we ask the question what makes a protein promiscuous, *i.e.*, prone to non-specific interactions, and structurally disordered. We predict that multi-scale correlations of amino acid positions within protein sequences statistically enhance the propensity for promiscuous intra- and inter-protein binding. We show that sequence correlations between amino acids of the same type are statistically enhanced in structurally disordered proteins and in hubs of organismal proteomes. We also show that structurally disordered proteins possess a significantly higher degree of sequence order than structurally ordered proteins. We develop an analytical theory for this effect and predict the robustness of our conclusions with respect to the amino acid composition and the form of the microscopic potential between the interacting sequences. Our findings have implications for understanding molecular mechanisms of protein aggregation diseases induced by the extension of sequence repeats.

# Introduction

Understanding molecular mechanisms providing specificity of protein-protein binding within a cell has been challenged by recent experimental evidences that a significant fraction of proteins in higher eukaryotes are either entirely or partially intrinsically disordered and thus each such protein presents an ensemble of structures [1; 2; 3; 4]. Introduction of high-throughput technologies for determining protein-protein interactions (PPI) enables researchers to address the key question: How molecular properties of individual proteins shape their global, physical interaction properties within a cell [5; 6]? The two methods that provide the dominant amount of such PPI data are the high-throughput yeast two-hybrid (Y2H) [7; 8; 9] and affinity purification followed by mass spectrometry (AP/MS) [10; 11; 12]. While the accuracy assessment of these experiments is still a matter for research [10; 13; 14], and they do not provide dynamical and functional properties of interactions [9], yet they do provide a remarkable snapshot of physical interaction map that might exist within a living cell. These experiments demonstrate that entire organismal proteomes possess a high degree of multi-specificity [9; 11; 12; 15; 16; 17], with a fraction of proteins (termed 'hubs') physically interacting with tens and even hundreds of partners. One open question is how these physical, binary PPI maps are related to functional, biological PPI maps [9; 18]? A closely related question is how the functional multi-specificity is linked to non-functional promiscuity [4; 19; 20; 21]? Are functionally multi-specific proteins inherently more promiscuous? In this paper we use the term 'promiscuity' or 'inter-protein promiscuity' to describe the propensity for enhanced non-specific binding between proteins. The term 'intra-protein promiscuity' is used here to describe the propensity for enhanced non-specific intra-molecular binding, which we suggest leads to an enhanced level of structural disorder.

In this paper we predict that protein sequences with enhanced correlations of sequence positions of amino acids of the same type generally represent more structurally disordered and more promiscuous sequences. In particular, we show that sequences of structurally disordered proteins possess significantly stronger correlations than sequences of structurally ordered proteins, such as all-alpha and all-beta proteins. We also show a strong signature of the predicted effect in hub proteins (identified by high-throughput Y2H and AP/MS experiments) of eukaryotic organismal proteomes.

Intuitively, sequence 'correlations' mean statistically significant repeats of sequence patterns. The existence of periodicities in protein sequences has been known since seminal works of Eisenberg *et al* [22; 23; 24; 25; 26; 27]. Most of these investigations show that sequence periodicities arise due to the existence of symmetrical structural elements such as alpha helices and beta sheets. More recently sequence periodicities have been also observed in disordered proteins [28; 29; 30; 31; 32]. It is also predicted that hub proteins possess a higher level of intrinsic disorder compared to non-hub proteins [33; 34]. Here we suggest a mechanism that explains why this is the case, and predicts which sequence correlations induce structural disorder and enhance promiscuity.

## Results

**Computational analysis of sequence correlations**

We first analyze the non-redundant dataset of structurally disordered proteins [35] (Materials and Methods). The comparison of correlation properties of structurally disordered proteins with all-alpha proteins shows that sequences of disordered proteins are significantly stronger correlated than sequences of perfectly structured proteins. We characterize the correlation properties of amino acid positions within protein sequences by the normalized correlation function,

$$\eta_{\alpha\beta}(x) = g_{\alpha\beta}(x) / \langle g^r_{\alpha\beta}(x) \rangle ,  \quad \text{Eq. (1)}$$

where $g_{\alpha\beta}(x)$ is the probability to find a residue of type $\alpha$ separated by the distance $x$ from a residue of type $\beta$, and $\langle g^r_{\alpha\beta}(x) \rangle$ is the corresponding probability for the randomized sequence set, averaged with respect to different random realizations (see Materials and Methods). If $\eta_{\alpha\beta}(x) > 1$, then the two residues $\alpha$ and $\beta$ are statistically correlated at the distance $x$; while for entirely random sequences, $\eta_{\alpha\beta}(x) = 1$. We emphasize that the definition, Eq. (1), removes the average amino acid compositional bias, and thus describes the correlation properties of different amino acid types on the same footing, despite the compositional differences between amino acids.

The twenty diagonal elements, $\eta_{\alpha\alpha}(x)$, of the entire correlation matrix, $\eta_{\alpha\beta}(x)$, for both disordered and all-alpha proteins datasets are represented in Figure 1A. The strength of diagonal correlations is significantly higher in disordered proteins than in all-alpha proteins (*p*-values are given in Table 1, see also Materials and Methods). The comparison of disordered proteins with all-beta proteins shows similar results (Supporting Information, Figure S1). The notable feature of the observed highly correlated sequence motifs is that strong correlations are observed at multiple length-scales, with the range of correlations reaching tens and even hundreds of residues. One interesting residue demonstrating strong, long-range correlations is Gly. In particular, there are 78 non-redundant proteins with a large number of repeats (contributing to the peaks in the correlation function, $\eta_{GG}(x)$) containing Gly in each of these proteins. There are 207 different Pfam domains [36] that constitute these 78 proteins. However, the most common domain, collagen, is shared only by four different proteins out of 78, Table S1. This shows that the observed effect is quite general and not dominated by one particular protein family.

To characterize the sequence correlations further for each group of proteins, we introduce the cumulative correlation function, $H_{\alpha\beta}$,

$$H_{\alpha\beta} = \sum_{x_i \in \max[\eta_{\alpha\beta}(x_i)]} \eta_{\alpha\beta}(x_i), \qquad \text{Eq. (2)}$$

where the summation is performed at the peaks of $\eta_{\alpha\beta}(x)$. Due to the fact that for many amino acids $\eta_{\alpha\beta}(x)$ exhibit enhanced correlations at different length-scales ($\eta_{GG}(x)$ represents the most striking example of such behavior, Figure 1), we have chosen to sum with respect to peaks in Eq. (2). This procedure takes into account the enhanced sequence correlations at all length-scales on the same footing, and not only short-range correlations (Materials and Methods). The ratio, $\chi_{\alpha\beta} = H_{\alpha\beta}^{dis} / H_{\alpha\beta}^{alpha}$, characterizes the relative correlation strength in disordered proteins compared to all-alpha proteins, Figure 1B and C (Figure S1, for all-beta proteins). A key observation here is that the diagonal correlations, $\chi_{\alpha\alpha}$, are the strongest. In particular, in disordered proteins the diagonal cumulative correlations are statistically significantly stronger than in all-alpha proteins, $\chi_{\alpha\alpha} > 1.1$, for G, Y, R, W, S, E, P, D, Q, A, K, T, and there no amino acids having an opposite trend (see Table 1 for $p$-values). When we remove 10% of longest sequences from our non-redundant dataset, ten out of twelve overrepresented amino acids remain significantly correlated; with only $\chi_{KK}$ slightly reduced, but still significant, $\chi_{KK} > 1.05$, and W is rejected by the $p$-value analysis, Figure S2. Our analytical model presented below predicts that the observed enhanced strength of diagonal correlations provides the enhanced promiscuity to the sequences.

We note that the computed set of strongly correlated amino acids linked to structural disorder: G, Y, R, W, S, E, P, D, Q, A, K, T, stems from a different origin than the known set of amino acids: P, Q, S, E, K, R, (and G is reported in some studies [37]), compositionally overrepresented (on average) in disordered regions of proteins [37]. Strikingly, we observe that compositionally low abundant amino acids, Y and W, show statistically significant correlation signal in disordered proteins.

In order to estimate the effect of sequence correlations on inter-protein promiscuity, we compare the correlation properties of hubs as compared to 'ends' (proteins having just one detectable interaction partner), in human, yeast and *E. coli*, Figure 2A. We used the high-quality Y2H PPI dataset produced by the Vidal group for the human proteome [9], and curated AP/MS datasets for yeast [11; 12; 15], and *E. coli* [16] (Materials and Methods). These high-throughput methods produce a snapshot of physical interactions on an entire proteome level. The key working hypothesis that we advocate here is that highly connected proteins possess the enhanced propensity for non-specific binding.

For six amino acids, H, F, I, P, G, Y, the diagonal cumulative sequence correlations in human hubs are statistically significantly stronger than in ends, $\chi_{\alpha\alpha} > 1.1$ (see Table 1 for *p*-values). Only for Q correlations are stronger in ends than in hubs, and for C and W the results are not statistically significant. Qualitatively similar, yet quantitatively weaker effect is observed in yeast, with W and Q, having stronger diagonal correlations in hubs than in ends ($\chi_{WW}, \chi_{QQ} > 1.1$), and one amino acid, H, having an opposite trend, $\chi_{HH} < 0.9$, Table 1. It is remarkable that $\chi_{HH}$ and $\chi_{QQ}$ behave qualitatively differently in human and yeast proteomes. Note that if we low the threshold to $\chi_{\alpha\alpha} > 1.05$, stronger diagonal correlations in hubs are observed for seven amino acids in yeast, W, Q, P, E, Y, A, and R. In *E.coli* the effect is yet weaker with only one amino acid, C, having stronger correlations in hubs, $\chi_{CC} > 1.1$, and if we low the threshold to $\chi_{\alpha\alpha} > 1.05$, statistically significant correlations are observed for C and T (see Table 1 for *p*-values). A notable feature observed in organismal interactomes is that P and Y demonstrate strong correlation effect for hubs both in human and yeast.

We also compared the correlation properties of the entire proteomes in *E. coli*, yeast and human, Figure 2B and C (Figure S3). For nine amino acids, P, H, K, E, R, S, G, A, and Y the diagonal correlations are stronger in the human proteome than in bacteria, $\chi_{\alpha\alpha} > 1.1$, with no amino acids showing an opposite trend. The computed *p*-values for all amino acids are significant, $p \leq 0.01$ (Materials and Methods). Our findings suggest therefore that human and yeast proteomes possess a statistically higher level of protein promiscuity compared to bacteria. This latter observation is also in agreement with a known prediction that the bacterial proteome contains practically no disordered proteins [3].

We predict below that enhanced diagonal sequence correlations observed in hub proteins lead to enhanced propensity for non-specific binding. Since many hub proteins detected in Y2H and AP/MS experiments are confirmed to be functionally multi-specific *in vivo* [9; 12; 18], our prediction suggests that enhanced non-specific binding and functional multi-specificity might be tightly linked. We suggest that a significant fraction of functionally multi-specific proteins might be inherently highly promiscuous.

**Theoretical model**

To answer the question why do enhanced diagonal sequence correlations lead to enhanced intra- and inter-protein promiscuity, we developed a simplified biophysical model that captures the effect of sequence correlations on the sequence interaction properties [38]. Here we provide an intuitive summary of our rigorous, analytical results [38]. The predicted effect originates from the enhanced symmetry of correlated sequences, and it is conceptually similar to the effect induced by the enhanced structural symmetry of interacting proteins investigated earlier [39; 40; 41; 42]. Formally promiscuity means that the free energy spectrum for non-specific binding is shifted towards lower energies. Specifically, we can define the probability distribution, $P(E)$, of the interaction energy, $E_A$, between the sequence $A$ with given correlation properties and a sequence from the target set (*e.g.* the set of random binders to sequence $A$). The sequence $A$ is designed using the stochastic procedure described below. The target set is not supposed to be optimized in any way for stronger binding with $A$. We can now compare the promiscuity of two sequences, $A$ and $B$ by comparing the standard deviations of the corresponding probability distributions, $\sigma_A$ and $\sigma_B$. We have recently shown analytically that if $\sigma_A > \sigma_B$, the corresponding free energies will always obey, $F_A < F_B$ [43], where $F_A$ and $F_B$ is the average interaction free energy of protein $A$ and $B$, respectively, with the target dataset. The averaging is performed with respect to different realizations of sequences $A$ and $B$, respectively, and with respect to different realizations of sequences from the target set. We assume that the average interaction energies are equal, $\langle E_A \rangle = \langle E_B \rangle$. The latter constraint is satisfied when the average amino acid compositions of $A$ and $B$ are equal [38]. We use this definition of promiscuity to describe both intra-protein and inter-protein promiscuity.

We first introduce a toy, one-dimensional lattice model that intuitively explains the nature of the predicted effect. We consider a set of sequences, each sequence of length $L$, with two types of amino acids, H and P, distributed at random positions along the sequence, Figure 3A. The amino acid types, H and P are entirely arbitrary. We now define the interaction energy, $E$, between a sequence from this set, and some 'target' sequence, $E = J \sum_{i=1}^{L} s_i$, where $s_i$ is a random variable that can acquire two values, $s_i = 1$ and $s_i = -1$, that corresponds to H and P amino acid types, respectively, at position $i$ along the protein sequence, and $J$ defines the inter-sequence binding strength. The averaging with respect to different sequence realizations obviously gives, $\langle s_i \rangle = 0$, $\langle s_i^2 \rangle = 1$, and $\langle s_i s_j \rangle = 0$, for any $i \neq j$. Therefore, the variance of $E$, $\sigma^2 = \langle E^2 \rangle = L \cdot J^2$. Next we assume that the sequence set is not random, but rather amino acids of the same type form adjacent pairs within the sequence, but there are no other correlations beyond the formed pairs, Figure 3B. In this case (we term it *C*2), due to the symmetry, the interaction energy between a pair of sequences, $E_{C2} = 2J \sum_{i=1}^{L/2} s_i$, and its variance, $\langle E_{C2}^2 \rangle = 2L \cdot J^2 = 2\sigma^2$, is twice as large as in the first case. In the case where three adjacent amino acids of the same type are clustered (*C*3), Figure 3C, the interaction energy, $E_{C3} = 3J \sum_{i=1}^{L/3} s_i$, and its variance, $\langle E_{C3}^2 \rangle = 3L \cdot J^2 = 3\sigma^2$. In all three cases the average energy, $\langle E \rangle$ is zero, and the only difference arises from the standard deviation of the probability distribution, $P(E)$, which is the Gaussian distribution according to the central limit theorem [44]. The longer is the correlation length of such homo-oligomer clusters, the higher the sequence symmetry, and the larger the standard deviation of $P(E)$. We have recently shown analytically that if two Gaussian probability distributions, $P(E_1)$ and $P(E_2)$, are characterized by non-equal standard deviations, $\sigma_1 > \sigma_2$, then the corresponding average free energies, $F_1$ and $F_2$, computed from these distributions always obey, $F_1 < F_2$ [43]. In particular, $F_1 - F_2 = -(\sigma_1^2 - \sigma_2^2)/2k_B T$, where $k_B$ is the Boltzmann constant, $T$ is the absolute temperature, $\sigma_1^2 = \langle E_1^2 \rangle$, and $\sigma_2^2 = \langle E_2^2 \rangle$ [43]. The key point here is that this result is invariant with respect to the sign of the inter-sequence binding constant, $J$. The strength of the predicted effect is entirely governed by the symmetry properties of sequence correlations. The toy model is easily generalizable to any number of amino acid types, simply by increasing the number of states in $s_i$, and leads to identical conclusions.

An example of a human protein sequence demonstrating strong diagonal correlations is shown in Fig. 3D. The Ewing sarcoma related protein (EWSR1) has 94 interaction partners, and is one of the strongest hubs in the human PPI network measured by the Vidal group [9]. Mutations in this oncogenic protein cause the Ewing sarcoma, a very aggressive, rare cancer occurring predominantly in children [45].

We introduce now a more detailed model for 'random' and 'designed' (or 'correlated') protein-like, sequences. Here we consider generic, long-range inter-sequence and intra-sequence interaction potentials. We also introduce a stochastic sequence 'design' procedure where we generate sequences with controllable strength and symmetry of sequence correlations. For simplicity we again use a minimalistic sequence alphabet with two types of amino acids, H and P. The notion of H and P amino acids stands here just in order to distinguish between two different amino acid types, and it does not constrain our conclusions to just hydrophobic and polar types. Our conclusions hold true for any number of amino acid types. Random sequence is obtained by distributing $N_h$ and $N_p$ amino acids at random positions within the one-dimensional sequence of the total length, $L = N_p + N_h$. Our simplistic approach therefore does not explicitly take into account the folding of the sequence. In order to obtain a correlated sequence, we allow residues to anneal at a given 'design' temperature, $T_d$. We impose that amino acids within the sequence under the design procedure interact through the pairwise additive design potential, $U_{\alpha\beta}(x)$, where $U_{pp}(x)$, $U_{hh}(x)$, and $U_{hp}(x)$ is the interaction potential between PP, HH, and HP amino acid types, respectively, and $x$ is the distance along the protein sequence. Each sequence configuration is then assigned its Boltzmann weight and accepted with the standard Metropolis criterion [46]. The only two assumptions about the interaction potentials, $U_{\alpha\beta}(x)$, are that they are pairwise additive and have a finite range of action. Examples of correlation functions computed for designed and random sequences are shown in Figure 4A.

Our next step is to analyze the probability distribution $P(E)$ of the interaction energy, $E$, between the random and correlated sequences. Every pair of interacting sequences consists of one random and one correlated sequence superimposed in a parallel configuration. The inter-sequence interaction potentials, $V_{pp}(\rho)$, $V_{hh}(\rho)$, and $V_{hp}(\rho)$ can have an arbitrary form, where $\rho$ is the inter-sequence distance. The probability distribution, $P(E)$, is characterized by its mean, $\langle E \rangle$, and by the variance, $\sigma^2$. The larger $\sigma$, and thus the broader the distribution, $P(E)$, the more promiscuous the correlated sequences. The mean is independent of the design potential, $U_{\alpha\beta}(x)$, and the variance, $\sigma^2 = \langle (E - \langle E \rangle)^2 \rangle$, possesses two key properties [38]. First, the more negative the effective 'design' potential, $U(x) = U_{pp} + U_{hh} - 2U_{ph}$, the greater is $\sigma$. One concludes that in order to increase $\sigma$ one needs to design the sequences with enhanced correlations in the positions between amino acids of similar types. This means that correlated sequences where amino acids of similar type are clustered together will be the more promiscuous sequences. Second, such correlated sequences will interact statistically stronger (than non-correlated sequences) with any set of arbitrary sequences independently on the sign of the inter-sequence interaction potential, $V(\rho)$. This is in accordance with the conclusion about the invariance of $\langle E^2 \rangle$ with respect to the sign of $J$, obtained above for the toy model. The predicted effects therefore are generic and qualitatively independent of the specific form of the inter-residue interaction potentials and on the amino acid composition of sequences. Intuitively, stronger correlations correspond to repetitive sequence patterns with a longer correlation length. The properties of the correlated patterns depend critically on the sign of the interaction potentials, $U_{\alpha\beta}(x)$, used in the design procedure. If the effective design potential $U(x)$ is overall negative (this corresponds to the attraction between the amino acids of similar types), the correlated patterns will have the form of repetitive residues of the same type, for example: HHHHPPPPHHHPPP... For entirely uncorrelated (random) sequences, all matrix elements of $\eta_{\alpha\beta}(x)$ are equal to unity. Clustering of amino acids of similar types corresponds to $\eta_{\alpha\alpha}(x) > 1$, Figure 4A. An example of correlation functions computed for the set of disordered proteins, where amino acids of similar types are statistically clustered is shown in Figure 4A. We computed $P(E)$ for two representative cases, Figure 4B. In the first case each interacting pair consists of designed and random sequences, and in the second case it consists of two random sequences, Figure 4B. In accordance with the analytical prediction, the standard deviation of $E$ in the first case is larger than the standard deviation in the second case, Figure 4B. The key message here is that enhanced correlations between amino acids of the same type lead to the broadening of the distribution, $P(E)$. Such broadening implies that the corresponding free energy will be always lower for stronger correlated sequences [43].

We stress that the presented simplified model does not explicitly take into account protein folding and, therefore, underestimates the effect of longer-range sequence correlations induced by the presence of a protein chain. Taking protein folding into account properly should provide an additional insight into the effect of long-range sequence correlations on protein structural disorder. Elucidation of the latter issue is the subject of our future work. We note that many disordered protein regions are predicted to contain linear sequence motifs that participate in numerous functional interactions [47]. It was recently suggested that such linear motifs play a key role in the dosage-induced toxicity effect [48]. Our theory is directly applicable to the latter case.

## Discussion

We note that consistent with our predictions, it was observed recently in [32] that homo-oligomer repeats are overrepresented in human, chimp, mouse, rat, and chicken genomes.

Remarkably, nine highly correlated residues observed in our analysis of disordered proteins: G, R, S, E, P, D, Q, A, and T, also appear among the residues forming frequent homo-oligomer repeats consisting of four repeated amino acids observed in [32]. In addition, we observed that three other amino acid types, Y, W, and K, possess strong diagonal correlations. Here we showed that this effect is much more general: It is the statistical propensity for promiscuous binding that is controlled by the strength of diagonal sequence correlations. One interesting example of highly correlated protein is titin, 2MDa_1, Figures S4 and S5. This long protein with 18,534 amino acids contains 41 repeated immunoglobulin I-set domains and 9 fibronectin Fn3 domains. However, the correlation analysis shows that the strongest correlations come from the 6000 amino acids long inter-domain region, Figure S4, S5. This region does not contain any immunoglobulin and fibronectin domains, but yet shows the strongest diagonal correlations. It is remarkable that the dominant contribution to diagonal correlations does not come from repeated structural domains. Based on our statistical analysis, this latter observation seems to represent the general rule.

We stress that a key prediction of the model is that highly promiscuous sequences (*i.e.* sequences with strong propensity to non-specific binding) possess strong diagonal correlations, $\eta_{\alpha\alpha}(x) > 1$, and at the same time, weak off-diagonal correlations, $\eta_{\alpha\beta}(x) < 1$, Figure 4A. Remarkably, we observe qualitatively similar behavior in the off-diagonal correlation functions of disordered proteins, Figure 5. In the examples presented in Figure 5 we show that the off-diagonal correlation functions are significantly reduced, $\eta_{\alpha\beta}(x) < 1$, for amino acids demonstrating strong diagonal correlations, such as Q, S, P, E, R, and A. Intuitively, strong diagonal correlations correspond to homo-oligomer repeats within protein sequences. Such repeats represent the most promiscuous sequences according to our model. It is known that many proteins involved in neurodegenerative diseases are mutated in a way that expands the length of repeated sequence regions. One of the most prominent examples is the huntingtin protein involved in Huntington's disease [49]. This protein has numerous repeats such as polyQ and polyP and it is known that the extended polyQ mutants are prone to aggregation [49].

In summary, here we suggest a mechanism that explains why sequence repeats of particular symmetries lead to enhanced protein promiscuity. We predict that both enhanced structural disorder and enhanced non-specific binding arise due to the enhancement of diagonal correlations of amino acid positions within protein sequences. We suggest that such enhanced diagonal correlations generically widen the energy spectrum of non-specific states within or between the proteins, which generically leads to the lowering of the free energy for disordered conformations or non-specific binding. In the former case (intra-protein promiscuity), the effect leads to the increased probability for thermodynamically allowed, non-native conformational states. We emphasize that our theoretical predictions concerning the intra-protein binding require further analysis, to take into account the actual folding of proteins with correlated sequences. However, a more detailed preliminary analysis based on theory of protein-like heteropolymers [50] indicates that enhanced sequence correlations give rise to greater structural flexibility of folded proteins. In the latter case (inter-protein promiscuity), this leads to the increased probability for thermodynamically allowed, non-specific binding states. This statistical effect is driven by symmetry properties of sequences and hence it is qualitatively robust with respect to parameters of the system, such as amino acid composition and specific form of the interaction potentials.

Our model description of inter- and intra-protein interactions is highly simplified, yet we suggest that the biophysical mechanism providing enhanced protein promiscuity described here is quite general and is likely to be the rule rather than the exception. The fact that many hub proteins detected by high-throughput Y2H and AP/MS screens are also confirmed being functionally multi-specific proteins [9; 12; 18] suggests that functional multi-specificity and non-functional promiscuity might be linked by a common design principle predicted here.

## Materials and Methods

**Disordered proteins dataset**

To compute sequence correlations in disordered proteins, we used the database of disordered proteins DisProt 5.4 [35], http://www.disprot.org/ . From this database, we selected a set of 547 non-redundant proteins with a mutual, pairwise sequence identity of less than 40%, Table S2. The experimentally known structurally ordered parts (such parts are systematically annotated in [35]) were removed from the selected set of sequences. We computed sequence correlations without aligning the sequences. In order to remove the bias, we did not include in our statistical correlation analysis the longest protein, titin (2MDa_1), with the length of 18,534 amino acids. We analyzed the correlation properties of this protein separately, Figures S4 and S5. We also performed an additional statistical significance test, removing from our curated dataset 10% of the longest sequences, and obtaining similar results, Figure S2. We computed the *p*-value for each reported value of $\chi_{\alpha\alpha}$, and the results are shown in Table 1.

**All-alpha and all-beta proteins dataset**

We selected the non-redundant sequences of all-alpha and all-beta proteins (with a pairwise sequence identity of less than 40%) from the Astral database [51], http://astral.berkeley.edu/ . We used a set of 1531 all alpha and 1348 all beta proteins, respectively, with the total sequence length equal approximately to the total sequence length of the disordered protein set.

**Hubs and ends in human, yeast, and bacterial proteomes**

**Human:** We used the yeast two-hybrid (Y2H), high-quality PPI data produced by the Vidal group [9]. Hubs were defined as proteins having 10 or more binding partners. Ends were defined as proteins for which only one interaction was detected. There were total of 107 hubs and 136 ends, Table S3. **Yeast:** We used the curated PPI dataset from [15] with a similar definition of hubs and ends as in the human PPI dataset. There were total 590 hubs and 693 ends, Table S4. ***E. coli*:** We used the PPI dataset from [16] for the K-12 strain W3110 with 277 hubs and 294 ends, Table S5. In each of the three organisms the total sequence length of hubs was approximately equal to the total sequence length of ends. In each organism the mutual, pairwise sequence identity in hubs and ends was imposed to be less than 75%.

**Entire proteome data**

The entire proteome sequences for human, yeast and *E. coli* (strain K-12, W3110) were downloaded from the NCBI site, http://www.ncbi.nlm.nih.gov/guide/genomes-maps/ . Each proteome was filtered from redundant proteins to reach the mutual, pairwise sequence identity of less than 40%. This left us with 12,033, 5112, and 3683 proteins in human, yeast, and *E. coli*, respectively.

**Analysis of sequence correlations**

In order to compute the average, $\langle g^r_{\alpha\beta}(x) \rangle$, in Eq. (1), we used five randomized replicas for each protein sequence. All presented correlation functions, $\eta_{\alpha\beta}(x)$, are computed with respect to the entire set of non-aligned protein sequences in a given group (*e.g.* disordered proteins, hubs, ends, entire proteomes for each organism, *etc.*). In order to compute the cumulative correlation functions, $H_{\alpha\beta}$, we have performed a summation with respect to five largest peaks of $\eta_{\alpha\beta}(x)$ in Eq. (2). We have also tried different definitions for the cumulative correlation function (for example, summation with respect to the first few consecutive distances, $x$, in Eq. (2)), which do not alter our qualitative conclusions, yet lead to quantitative differences (data not shown). The summation with respect to peaks of $\eta_{\alpha\beta}(x)$ has the advantage that it takes into account the strongest sequence correlations at all length-scales on the same footing (and not only short-range correlations). In our estimates of amino acids with significantly overrepresented correlation strength, we have chosen to use 10% or 5% threshold for the relative correlation strength, $\chi_{\alpha\alpha}$. If thus $\chi_{\alpha\alpha} > 1.1$ or $\chi_{\alpha\alpha} > 1.05$, and the corresponding *p*-value, $p < 0.05$, we assigned $\alpha$ to a set of significantly correlated amino acids.

***p*-value calculations**

In order to compute the *p*-values, we first prepared one hundred reshuffled replicas for each sequence dataset (i.e., for disordered proteins; for hubs and ends in human, yeast and *E. coli;* and for the entire proteomes of these organisms, as explained above). We reshuffled each protein sequence separately, without changing its average amino acid composition, and just exchanging the sequence positions of amino acids. In order to estimate the *p*-value, we performed one thousand calculations for each case, exactly as described in the main text, but every time using the corresponding randomized dataset instead of actual, biological sequences. We then counted the number of occurrences for the randomized dataset, $N_{rand}(\chi_{\alpha\beta}^{rand} \geq \chi_{\alpha\beta})$, when, $\chi_{\alpha\beta}^{rand} \geq \chi_{\alpha\beta}$, and estimated the corresponding *p*-value, $p \simeq N_{rand}/1000$. Due to a high computational complexity for entire proteome *p*-value calculations, we assigned statistical significance using only hundred randomized datasets.

## Acknowledgements

We thank Amir Aharoni, Gilad Haran, Nikolaus Rajewsky, Irit Sagi, Itamar Sela, Dan Tawfik, and the members of Marc Vidal's lab for helpful discussions. D. L. acknowledges the financial support from the Israel Science Foundation grant 1014/09. A. A. is a recipient of the Lewiner graduate fellowship.

## References

1. Dyson, H. J. & Wright, P. E. (2005). Intrinsically unstructured proteins and their functions. *Nat Rev Mol Cell Biol* **6**, 197-208.
2. Wright, P. E. & Dyson, H. J. (2009). Linking folding and binding. *Curr Opin Struct Biol* **19**, 31-8.
3. Dunker, A. K., Silman, I., Uversky, V. N. & Sussman, J. L. (2008). Function and structure of inherently disordered proteins. *Curr Opin Struct Biol* **18**, 756-64.
4. Zhang, J., Maslov, S. & Shakhnovich, E. I. (2008). Constraints imposed by non-functional protein-protein interactions on gene expression and proteome size. *Mol Syst Biol* **4**, 210.
5. Aloy, P. & Russell, R. B. (2006). Structural systems biology: modelling protein interactions. *Nat Rev Mol Cell Biol* **7**, 188-97.
6. Kim, P. M., Lu, L. J., Xia, Y. & Gerstein, M. B. (2006). Relating three-dimensional structures to protein networks provides evolutionary insights. *Science* **314**, 1938-41.
7. Ito, T., Chiba, T., Ozawa, R., Yoshida, M., Hattori, M. & Sakaki, Y. (2001). A comprehensive two-hybrid analysis to explore the yeast protein interactome. *Proc Natl Acad Sci U S A* **98**, 4569-74.
8. Uetz, P., Giot, L., Cagney, G., Mansfield, T. A., Judson, R. S., Knight, J. R., Lockshon, D., Narayan, V., Srinivasan, M., Pochart, P., Qureshi-Emili, A., Li, Y., Godwin, B., Conover, D., Kalbfleisch, T., Vijayadamodar, G., Yang, M., Johnston, M., Fields, S. & Rothberg, J. M. (2000). A comprehensive analysis of protein-protein interactions in Saccharomyces cerevisiae. *Nature* **403**, 623-7.
9. Rual, J. F., Venkatesan, K., Hao, T., Hirozane-Kishikawa, T., Dricot, A., Li, N., Berriz, G. F., Gibbons, F. D., Dreze, M., Ayivi-Guedehoussou, N., Klitgord, N., Simon, C., Boxem, M., Milstein, S., Rosenberg, J., Goldberg, D. S., Zhang, L. V., Wong, S. L., Franklin, G., Li, S., Albala, J. S., Lim, J., Fraughton, C., Llamosas, E., Cevik, S., Bex, C., Lamesch, P., Sikorski, R. S., Vandenhaute, J., Zoghbi, H. Y., Smolyar, A., Bosak, S., Sequerra, R., Doucette-Stamm, L., Cusick, M. E., Hill, D. E., Roth, F. P. & Vidal, M. (2005). Towards a proteome-scale map of the human protein-protein interaction network. *Nature* **437**, 1173-8.
10. Yu, H., Braun, P., Yildirim, M. A., Lemmens, I., Venkatesan, K., Sahalie, J., Hirozane-Kishikawa, T., Gebreab, F., Li, N., Simonis, N., Hao, T., Rual, J. F., Dricot, A., Vazquez, A., Murray, R. R., Simon, C., Tardivo, L., Tam, S., Svrzikapa, N., Fan, C., de Smet, A. S., Motyl, A., Hudson, M. E., Park, J., Xin, X., Cusick, M. E., Moore, T., Boone, C., Snyder, M., Roth, F. P., Barabasi, A. L., Tavernier, J., Hill, D. E. & Vidal, M. (2008). High-quality binary protein interaction map of the yeast interactome network. *Science* **322**, 104-10.
11. Gavin, A. C., Aloy, P., Grandi, P., Krause, R., Boesche, M., Marzioch, M., Rau, C., Jensen, L. J., Bastuck, S., Dumpelfeld, B., Edelmann, A., Heurtier, M. A., Hoffman, V., Hoefert, C., Klein, K., Hudak, M., Michon, A. M., Schelder, M., Schirle, M., Remor, M.,

Rudi, T., Hooper, S., Bauer, A., Bouwmeester, T., Casari, G., Drewes, G., Neubauer, G., Rick, J. M., Kuster, B., Bork, P., Russell, R. B. & Superti-Furga, G. (2006). Proteome survey reveals modularity of the yeast cell machinery. *Nature* **440**, 631-6.
12. Krogan, N. J., Cagney, G., Yu, H., Zhong, G., Guo, X., Ignatchenko, A., Li, J., Pu, S., Datta, N., Tikuisis, A. P., Punna, T., Peregrin-Alvarez, J. M., Shales, M., Zhang, X., Davey, M., Robinson, M. D., Paccanaro, A., Bray, J. E., Sheung, A., Beattie, B., Richards, D. P., Canadien, V., Lalev, A., Mena, F., Wong, P., Starostine, A., Canete, M. M., Vlasblom, J., Wu, S., Orsi, C., Collins, S. R., Chandran, S., Haw, R., Rilstone, J. J., Gandi, K., Thompson, N. J., Musso, G., St Onge, P., Ghanny, S., Lam, M. H., Butland, G., Altaf-Ul, A. M., Kanaya, S., Shilatifard, A., O'Shea, E., Weissman, J. S., Ingles, C. J., Hughes, T. R., Parkinson, J., Gerstein, M., Wodak, S. J., Emili, A. & Greenblatt, J. F. (2006). Global landscape of protein complexes in the yeast Saccharomyces cerevisiae. *Nature* **440**, 637-43.
13. Deeds, E. J., Ashenberg, O. & Shakhnovich, E. I. (2006). A simple physical model for scaling in protein-protein interaction networks. *Proc Natl Acad Sci U S A* **103**, 311-6.
14. Aloy, P. & Russell, R. B. (2002). Potential artefacts in protein-interaction networks. *FEBS Lett* **530**, 253-4.
15. Batada, N. N., Reguly, T., Breitkreutz, A., Boucher, L., Breitkreutz, B. J., Hurst, L. D. & Tyers, M. (2006). Stratus not altocumulus: a new view of the yeast protein interaction network. *PLoS Biol* **4**, e317.
16. Hu, P., Janga, S. C., Babu, M., Diaz-Mejia, J. J., Butland, G., Yang, W., Pogoutse, O., Guo, X., Phanse, S., Wong, P., Chandran, S., Christopoulos, C., Nazarians-Armavil, A., Nasseri, N. K., Musso, G., Ali, M., Nazemof, N., Eroukova, V., Golshani, A., Paccanaro, A., Greenblatt, J. F., Moreno-Hagelsieb, G. & Emili, A. (2009). Global functional atlas of Escherichia coli encompassing previously uncharacterized proteins. *PLoS Biol* **7**, e96.
17. Bork, P., Jensen, L. J., von Mering, C., Ramani, A. K., Lee, I. & Marcotte, E. M. (2004). Protein interaction networks from yeast to human. *Curr Opin Struct Biol* **14**, 292-9.
18. Han, J. D., Bertin, N., Hao, T., Goldberg, D. S., Berriz, G. F., Zhang, L. V., Dupuy, D., Walhout, A. J., Cusick, M. E., Roth, F. P. & Vidal, M. (2004). Evidence for dynamically organized modularity in the yeast protein-protein interaction network. *Nature* **430**, 88-93.
19. Khersonsky, O. & Tawfik, D. S. (2010). Enzyme promiscuity: a mechanistic and evolutionary perspective. *Annu Rev Biochem* **79**, 471-505.
20. Nobeli, I., Favia, A. D. & Thornton, J. M. (2009). Protein promiscuity and its implications for biotechnology. *Nat Biotechnol* **27**, 157-67.
21. Heo, M., Maslov, S. & Shakhnovich, E. I. (2011). Topology of protein inetraction network shapes protein abundances and their functional and non-specific interactions. *Proc Natl Acad Sci U S A* **doi:10.1073/pnas.1009392108**.
22. Eisenberg, D., Weiss, R. M. & Terwilliger, T. C. (1984). The hydrophobic moment detects periodicity in protein hydrophobicity. *Proc Natl Acad Sci U S A* **81**, 140-4.
23. Cornette, J. L., Margalit, H., Berzofsky, J. A. & DeLisi, C. (1995). Periodic variation in side-chain polarities of T-cell antigenic peptides correlates with their structure and activity. *Proc Natl Acad Sci U S A* **92**, 8368-72.
24. Marcotte, E. M., Pellegrini, M., Yeates, T. O. & Eisenberg, D. (1999). A census of protein repeats. *J Mol Biol* **293**, 151-60.

25. Rackovsky, S. (1998). "Hidden" sequence periodicities and protein architecture. *Proc Natl Acad Sci U S A* **95**, 8580-4.
26. Rosato, V., Pucello, N. & Giuliano, G. (2002). Evidence for cysteine clustering in thermophilic proteomes. *Trends Genet* **18**, 278-81.
27. Xiong, H., Buckwalter, B. L., Shieh, H. M. & Hecht, M. H. (1995). Periodicity of polar and nonpolar amino acids is the major determinant of secondary structure in self-assembling oligomeric peptides. *Proc Natl Acad Sci U S A* **92**, 6349-53.
28. Dosztanyi, Z., Chen, J., Dunker, A. K., Simon, I. & Tompa, P. (2006). Disorder and sequence repeats in hub proteins and their implications for network evolution. *J Proteome Res* **5**, 2985-95.
29. Jorda, J., Xue, B., Uversky, V. N. & Kajava, A. V. (2010). Protein tandem repeats - the more perfect, the less structured. *FEBS J* **277**, 2673-82.
30. Tompa, P. (2003). Intrinsically unstructured proteins evolve by repeat expansion. *Bioessays* **25**, 847-55.
31. Lise, S. & Jones, D. T. (2005). Sequence patterns associated with disordered regions in proteins. *Proteins* **58**, 144-50.
32. Simon, M. & Hancock, J. M. (2009). Tandem and cryptic amino acid repeats accumulate in disordered regions of proteins. *Genome Biol* **10**, R59.
33. Haynes, C., Oldfield, C. J., Ji, F., Klitgord, N., Cusick, M. E., Radivojac, P., Uversky, V. N., Vidal, M. & Iakoucheva, L. M. (2006). Intrinsic disorder is a common feature of hub proteins from four eukaryotic interactomes. *PLoS Comput Biol* **2**, e100.
34. Kim, P. M., Sboner, A., Xia, Y. & Gerstein, M. (2008). The role of disorder in interaction networks: a structural analysis. *Mol Syst Biol* **4**, 179.
35. Sickmeier, M., Hamilton, J. A., LeGall, T., Vacic, V., Cortese, M. S., Tantos, A., Szabo, B., Tompa, P., Chen, J., Uversky, V. N., Obradovic, Z. & Dunker, A. K. (2007). DisProt: the Database of Disordered Proteins. *Nucleic Acids Res* **35**, D786-93.
36. Finn, R. D., Mistry, J., Tate, J., Coggill, P., Heger, A., Pollington, J. E., Gavin, O. L., Gunasekaran, P., Ceric, G., Forslund, K., Holm, L., Sonnhammer, E. L., Eddy, S. R. & Bateman, A. (2010). The Pfam protein families database. *Nucleic Acids Res* **38**, D211-22.
37. He, B., Wang, K., Liu, Y., Xue, B., Uversky, V. N. & Dunker, A. K. (2009). Predicting intrinsic disorder in proteins: an overview. *Cell Res* **19**, 929-49.
38. Lukatsky, D. B. & Afek, A. Sequence correlations shape protein promiscuity. *arXiv:1004.5048v2*.
39. Lukatsky, D. B., Shakhnovich, B. E., Mintseris, J. & Shakhnovich, E. I. (2007). Structural similarity enhances interaction propensity of proteins. *J Mol Biol* **365**, 1596-606.
40. Lukatsky, D. B. & Shakhnovich, E. I. (2008). Statistically enhanced promiscuity of structurally correlated patterns. *Phys Rev E Stat Nonlin Soft Matter Phys* **77**, 020901.
41. Lukatsky, D. B., Zeldovich, K. B. & Shakhnovich, E. I. (2006). Statistically enhanced self-attraction of random patterns. *Phys Rev Lett* **97**, 178101.
42. Andre, I., Strauss, C. E., Kaplan, D. B., Bradley, P. & Baker, D. (2008). Emergence of symmetry in homooligomeric biological assemblies. *Proc Natl Acad Sci U S A* **105**, 16148-52.
43. Lukatsky, D. B. & Elkin, M. Energy fluctuations shape free energy of biomolecular interactions. *arXiv:1101.4529v1*.
44. Feller, W. (1970). *An introduction to probability theory and its applications*. 3rd edit, John Wiley & Sons, New York.

45. Erkizan, H. V., Uversky, V. N. & Toretsky, J. A. Oncogenic partnerships: EWS-FLI1 protein interactions initiate key pathways of Ewing's sarcoma. *Clin Cancer Res* **16**, 4077-83.
46. Frenkel, D. & Smit, B. (2002). *Understanding molecular simulation : from algorithms to applications*. 2nd edit, Academic Press, San Diego.
47. Russell, R. B. & Gibson, T. J. (2008). A careful disorderliness in the proteome: sites for interaction and targets for future therapies. *FEBS Lett* **582**, 1271-5.
48. Vavouri, T., Semple, J. I., Garcia-Verdugo, R. & Lehner, B. (2009). Intrinsic protein disorder and interaction promiscuity are widely associated with dosage sensitivity. *Cell* **138**, 198-208.
49. Cohen, E. & Dillin, A. (2008). The insulin paradox: aging, proteotoxicity and neurodegeneration. *Nat Rev Neurosci* **9**, 759-67.
50. Ramanathan, S. & Shakhnovich, E. (1994). Statistical mechanics of proteins with "evolutionary selected" sequences. *Phys Rev E Stat Phys Plasmas Fluids Relat Interdiscip Topics* **50**, 1303-1312.
51. Chandonia, J. M., Hon, G., Walker, N. S., Lo Conte, L., Koehl, P., Levitt, M. & Brenner, S. E. (2004). The ASTRAL Compendium in 2004. *Nucleic Acids Res* **32**, D189-92.

**Table Legends**

**Table 1.** *p*-values for statistical significance of cumulative correlation functions, $\chi_{\alpha\alpha}$, computed for all analyzed datasets. The details of *p*-value calculations are given in Materials and Methods. Zero values correspond to $p < 0.001$.

**Figure Legends**

**Figure 1. Sequence correlations in structurally disordered proteins compared to all-alpha proteins.**
**A.** Diagonal elements of the correlation function, $\eta_{\alpha\alpha}(x)$, for all 20 types of amino acids for intrinsically disordered protein (red circles) and all-alpha proteins (blue diamonds). **B.** The heat-map representation of the entire cumulative correlations matrix, $H_{\alpha\beta}$, for disordered, and all-alpha proteins. **C.** The diagonal elements of the ratio, $\chi_{\alpha\beta} = H_{\alpha\beta}^{dis} / H_{\alpha\beta}^{alpha}$.

**Figure 2. Sequence correlations in hubs, ends and entire proteomes in human, yeast, and *E. coli*.**
**A.** Relative correlation strength of hubs compared to ends, $\chi_{\alpha\alpha} = H_{\alpha\alpha}^{hubs} / H_{\alpha\alpha}^{ends}$, for human, yeast, and *E. coli*, respectively. We detected strong compositional variability of Cys in the dataset of human ends, making the result for $\chi_{CC}$ not statistically significant. **B.** The cumulative correlation matrix, $H_{\alpha\beta}$, for the entire human and *E. coli* proteomes, respectively. **C.** Relative correlation strength, $\chi_{\alpha\alpha} = H_{\alpha\alpha}^{human} / H_{\alpha\alpha}^{ecoli}$, of diagonal correlations of the entire human compared to the entire *E. coli* proteomes.

**Figure 3. Diagonal sequence correlations and promiscuity. Toy model.**
Cartoon for the toy model shows the pairs of interacting sequences, where one sequence has a varying symmetry of sequence correlations, random sequence (**A**), designed sequence with perfectly correlated nearest-neighbor amino acids of the same type (**B**), and designed sequence with perfectly correlated nearest-neighbor and next nearest-neighbor amino acids of the same type (**C**). We term such correlations 'diagonal'. The longer the correlation length of diagonal sequence correlations, the larger the standard deviation of the binding energy spectrum. (**D**) Example: One of the strongest hubs from the human PPI network [9], the Ewing sarcoma related protein EWSR1, also known as EWS. This 657 amino acids long protein has 94 interaction partners. Some of the nearest-neighbor correlated amino acids are marked with color. There are numerous additional higher-order diagonal correlations in this protein, which are not explicitly shown, but can be recognized by an eye. Mutations in this oncogenic protein cause the Ewing sarcoma, a very aggressive, rare bone cancer occurring predominantly in children.

**Figure 4. Sequence correlations and distribution of interaction energies in model sequences.**

**A.** Computed sequence correlation functions for the 'designed' sequences, $\eta_{pp}(x) = \eta_{hh}(x)$ (red squares), $\eta_{hp}(x) = \eta_{ph}(x)$ (blue diamonds); and for the random sequences, (black circles). The design potential was chosen to be $U_{pp} = U_{hh} = -1$, and $U_{hp} = 1$, and we assumed that only the nearest-neighbor amino acids can interact. The design temperature is $T_d = 2$. The sequence length is 200 amino acids, and we generated 5000 different sequences in each calculation. The uniform amino acid composition was adopted: 50% P and 50% H amino acids in each sequence. The error bars are smaller than the symbol size. Insert: Example: Diagonal, $\eta_{RR}(x)$, and off diagonal, $\eta_{RD}(x)$, correlation functions computed using the sequences of structurally disordered proteins (as in Figure 1A). **B**. Computed probability distribution function, $P(E)$, for the interaction energies between pairs of two random sequences (grey), and pairs consisting each of random and designed sequences, where designed sequences were generated at $T_d = 1$ (red). The energy $E$ is normalized per one amino acid.

**Figure 5. Example of off-diagonal correlation functions in structurally disordered proteins.**

Example: computed off-diagonal elements of the correlation functions, $\eta_{\alpha\beta}(x)$, for structurally disordered protein dataset. These correlation functions behave in a qualitative agreement with the behavior predicted for the off-diagonal elements of $\eta_{\alpha\beta}(x)$ in the model promiscuous sequences, Figure 4A, where $\eta_{\alpha\beta}(x) < 1$.

|   | Disprot/All alpha | Hubs/ Ends Human | Hubs/ Ends Yeast | Hubs/Ends E.coli |
|---|---|---|---|---|
| C | 0.017 | - | 0.431 | 0.001 |
| M | 0.188 | 0.09 | 0.301 | 0.13 |
| F | 0 | 0 | 0.047 | 0.901 |
| I | 0 | 0 | 0.162 | 0.043 |
| L | 0 | 0 | 0.008 | 0.025 |
| V | 0 | 0.358 | 0.297 | 0.03 |
| W | 0.004 | 0.143 | 0 | 0.695 |
| Y | 0 | 0.013 | 0.006 | 0.601 |
| A | 0 | 0.406 | 0 | 0.275 |
| G | 0 | 0 | 0 | 0.474 |
| T | 0 | 0.047 | 0.013 | 0.009 |
| S | 0 | 0.399 | 0 | 0.332 |
| N | 0.001 | 0.075 | 0.084 | 0.338 |
| Q | 0 | 0.017 | 0 | 0.021 |
| D | 0 | 0.365 | 0 | 0.076 |
| E | 0 | 0.07 | 0 | 0.858 |
| H | 0 | 0 | 0 | 0.052 |
| R | 0 | 0 | 0 | 0.744 |
| K | 0 | 0.07 | 0.001 | 0.596 |
| P | 0 | 0 | 0 | 0.038 |

**Table 1**

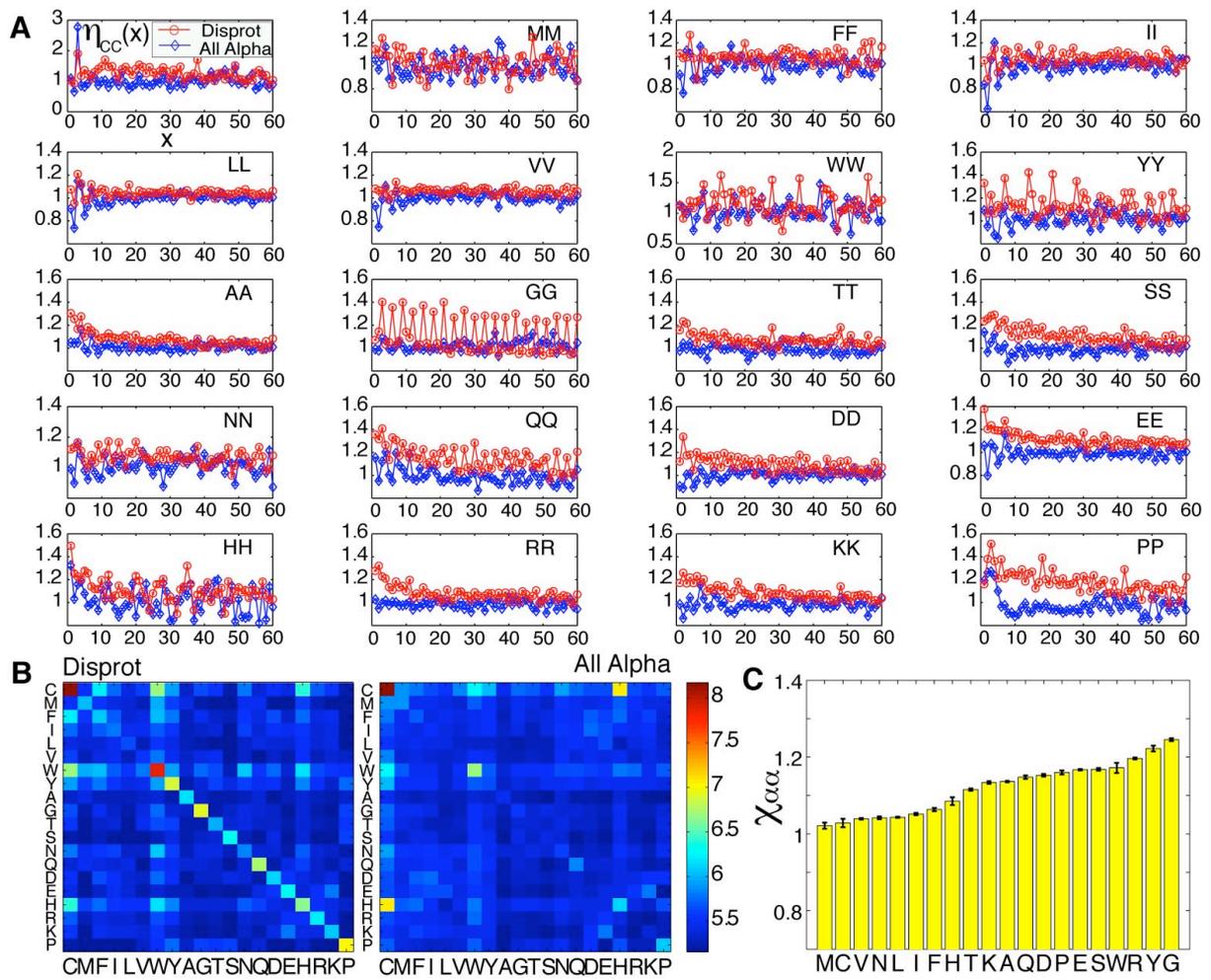

Figure 1

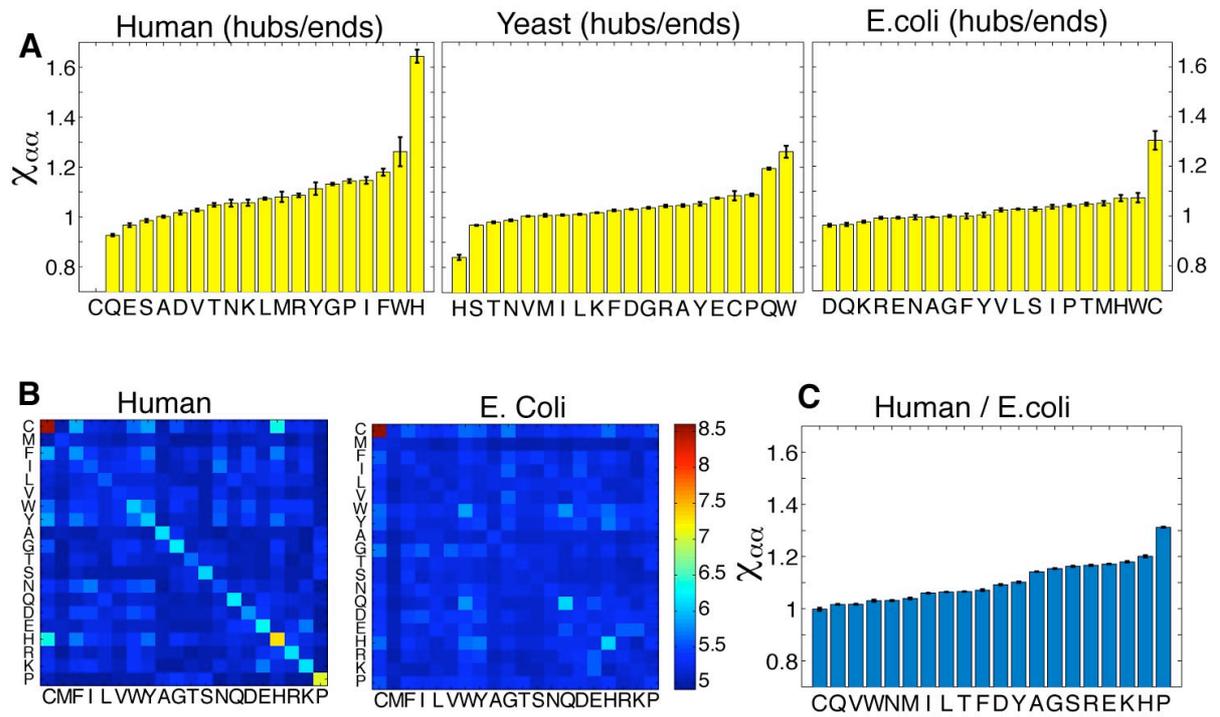

**Figure 2**

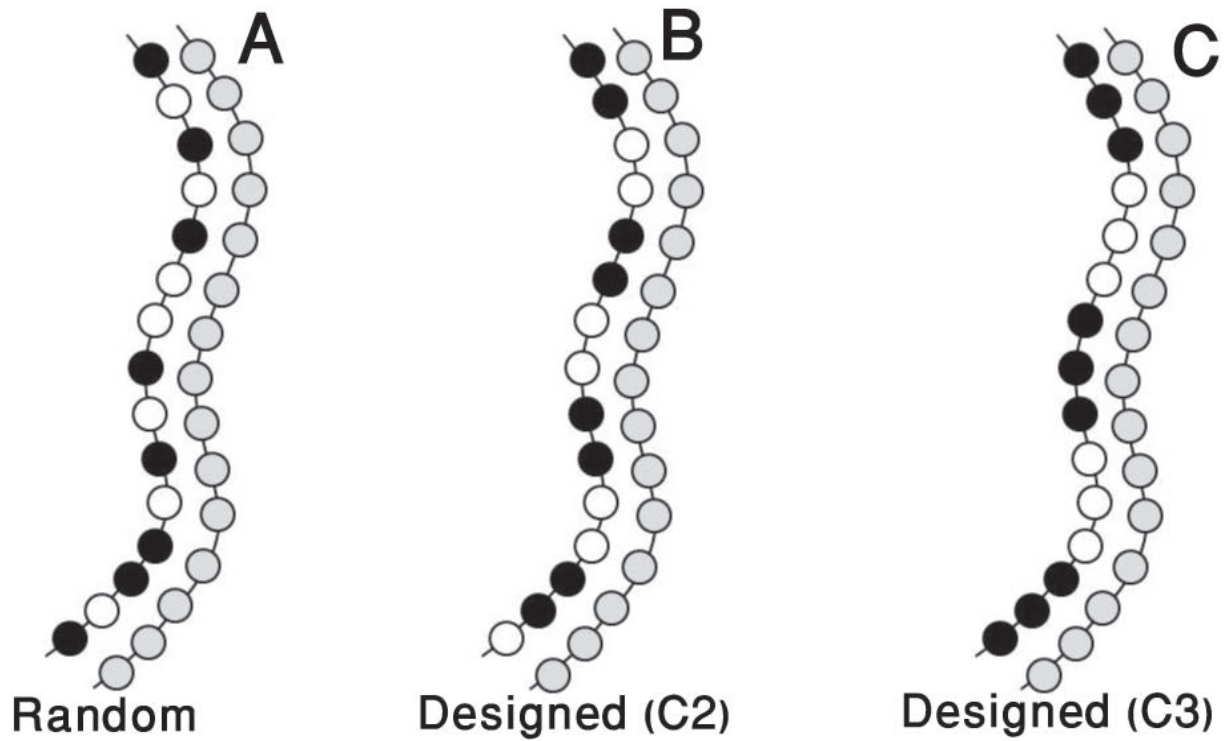

**Figure 3**

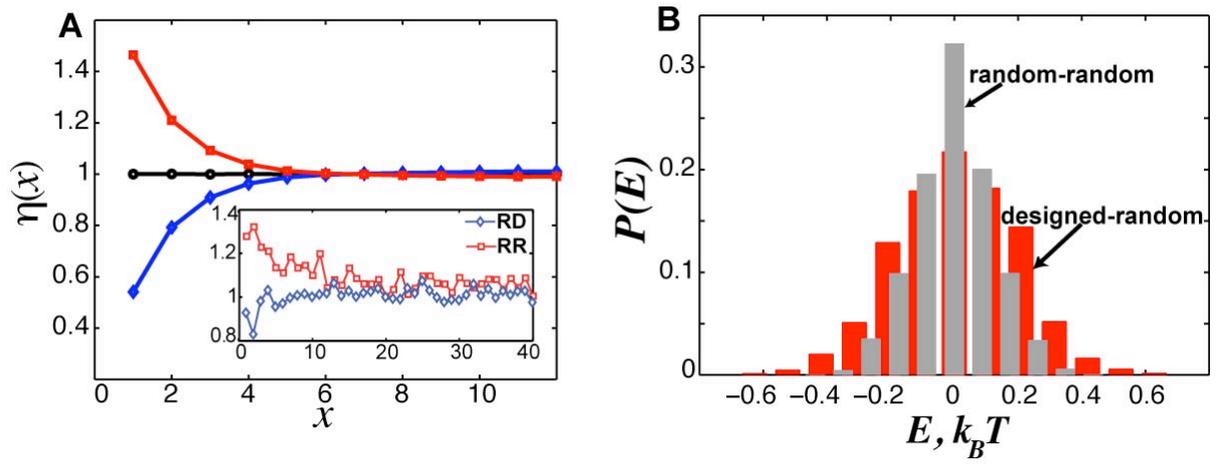

**Figure 4**

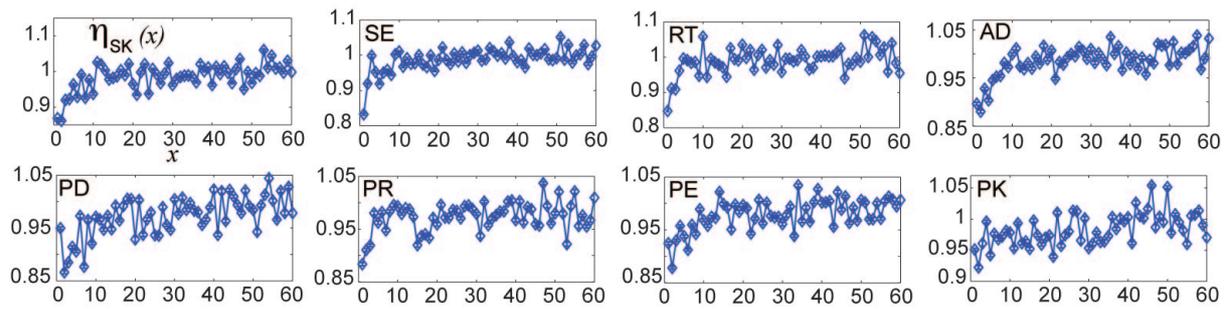

**Figure 5**

## Supplementary Data

**Supplementary data to this article can be found online at doi:10.1016/j.jmb.2011.03.056**

Ariel Afek, Eugene I. Shakhnovich, and David B. Lukatsky, "*Multi-scale sequence correlations increase proteome structural disorder and promiscuity*", Journal of Molecular Biology, **409** (2011) pp. 439-449.